\documentclass[twocolumn,english,notitlepage,twocolumn,showpacs,superscriptaddress]{revtex4-1}
\usepackage[T1]{fontenc}
\usepackage[latin9]{inputenc}
\usepackage{mathrsfs}
\usepackage{amsmath}
\usepackage{graphicx}

\makeatletter

\newcommand*\LyXThinSpace{\,\hspace{0pt}}
\providecommand{\tabularnewline}{\\}

\usepackage{babel}

\makeatother

\usepackage{babel}
\begin{document}

\title{Optimal operating protocol to achieve efficiency at maximum power
of heat engines}

\author{Yu-Han Ma}

\affiliation{Beijing Computational Science Research Center, Beijing 100193, China}

\affiliation{Graduate School of Chinese Academy of Engineering Physics, Beijing
100084, China}

\author{Dazhi Xu}
\email{dzxu@bit.edu.cn}

\selectlanguage{english}%

\affiliation{Center for Quantum Technology Research and School of Physics, Beijing
Institute of Technology, Beijing 100081, China}

\affiliation{Graduate School of Chinese Academy of Engineering Physics, Beijing
100084, China}

\author{Hui Dong}
\email{hdong@gscaep.ac.cn}

\selectlanguage{english}%

\affiliation{Graduate School of Chinese Academy of Engineering Physics, Beijing
100084, China}

\author{Chang-Pu Sun}

\affiliation{Beijing Computational Science Research Center, Beijing 100193, China}

\affiliation{Graduate School of Chinese Academy of Engineering Physics, Beijing
100084, China}
\begin{abstract}
The efficiency at maximum power has been investigated extensively,
yet the practical control scheme to achieve it remains elusive. We
fill such gap with a stepwise Carnot-like cycle, which consists the
discrete isothermal process (DIP) and adiabatic process. With DIP,
we validate the widely adopted assumption of $\mathscr{C}/t$ relation
of the irreversible entropy generation $S^{(\mathrm{ir})}$, and show
the explicit dependence of the coefficient $\mathscr{C}$ on the fluctuation
of the speed of tuning energy levels as well as the microscopic coupling
constants to the heat baths. Such dependence allows to control the
irreversible entropy generation by choosing specific control schemes.
We further demonstrate the achievable efficiency at maximum power
and the corresponding control scheme with the simple two-level system.
Our current work opens new avenues for the experimental test, which
was not feasible due to the lack the of the practical control scheme
in the previous low-dissipation model or its equivalents.
\end{abstract}
\maketitle

\section{Introduction}

Designing optimal heat engine is one of the primary goals in the recent
flourishing studies of heat engines both classically \cite{CA-endoreversible-heat-engine,actual-heat-engine,endoreversible-HEs,Linear-irreversible-heat-engine,Linear-irreversible-HE-1}
and quantum mechanically \cite{density-matrix-expansion-1,quantun,TLA-heat-engine}.
One of the most important characteristics is the output power, which
measures the energy output per unit of time. When the output power
achieves its maximum value, the corresponding efficiency, known as
efficiency at maximum power (EMP) \cite{EMP0,EMP1,EMP2,EMP-Tu,EMP3,constriant2-1,constriant2},
is another important characteristic of the heat engine. The achievable
EMP is well investigated via low-dissipation model \cite{low-dissipation},
which is recently proved to be equivalent to the linear response model
\cite{Linear-irreversible-low-disspation}. The low-dissipation model
simply assumes that the irreversible entropy generation, characterizing
the irreversibility, is inversely proportional to the operation time
$t$ with a coefficient $\mathscr{C}$, namely $\mathscr{C}/t$ relation.
The EMP is achieved via optimizing the operation times, as well as
the coefficients. However, such simple model leaves two major questions:
(1) how universal is the $\mathscr{C}/t$ relation? and (2) what is
the control protocol to achieve the corresponding EMP? The second
question is critical to the engine design, as well as the experimental
test.

The main obstacle to answer the two underline questions is the lack
of a microscopic model, with which the operating cycle can be shown
explicitly, and kept simple enough to allow analytically proof. To
maintain efficiency, it is meaningful to follow the Carnot cycle,
which consists two isothermal processes and two adiabatic processes.
The main difficulty is to design a quasi-isothermal process, which
refers to a process with finite operation time while in contact with
a heat bath. We have initialized such attempt to overcome the difficulty
in our previous work \cite{TLA-heat-engine}, yet limited to two-level
system with a simple linear tuning of energy levels.

In the current paper, we design a discrete isothermal process, which
consists a series of quantum isochoric and quantum adiabatic sub-processes.
With such discrete process, the $\mathscr{C}/t$ relation is analytically
validated in the low-dissipation region for arbitrary finite-dimension
system under arbitrary control scheme. Moreover, we obtain for the
first time the exact dependence of coefficient $\mathscr{C}$ on a
few parameters of the control scheme in the discrete isothermal process.
Based on this discovery, we design a two-level stepwise Carnot-like
heat engine and tune energy levels in different ways when it contacts
with the high temperature and low temperature heat bath. As a result,
the EMP of such heat engine is found to be controllable, and in some
circumstances, can be effectively improved.

\section{Irreversible entropy generation in discrete isothermal process}

In this section, we will construct a discrete process operating under
finite time to resemble the isothermal process in Carnot cycle and
prove the $\mathscr{C}/t$ relation. The Carnot efficiency, $\eta_{C}=1-T_{C}/T_{H}$,
is the fundamental upper bound, which a heat engine working between
two heat baths with temperatures $T_{H}$ and $T_{C}$ can achieve
\cite{Carnot}. Naturally, it is straightforward to design Carnot-like
process to achieve maximum efficiency under given output power. The
key question is how to realize a finite-time operation resembling
isothermal process, which usually takes infinite time in Carnot cycle.

\begin{figure}
\begin{centering}
\includegraphics[width=8.5cm]{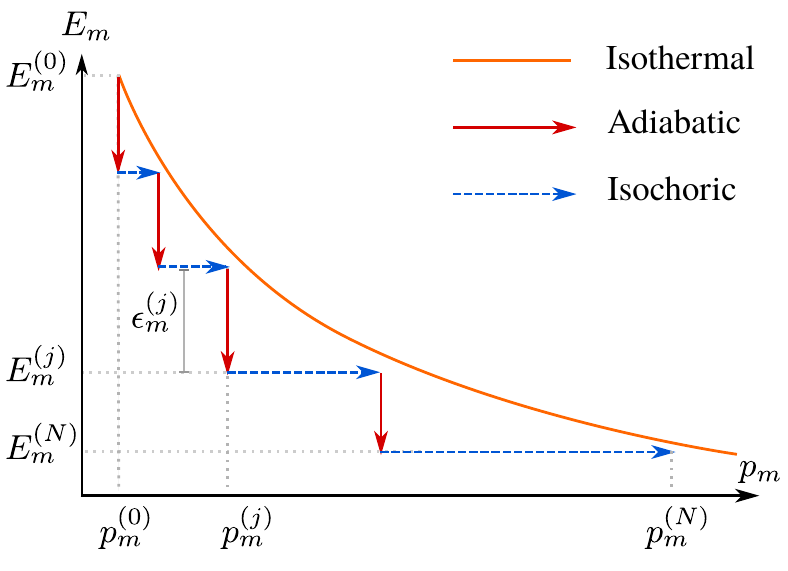} 
\par\end{centering}
\caption{\label{fig-discrete isothermal process-1}(Color online) Schematic
illustration of a discrete isothermal process (DIP). The horizontal
axis in this figure is the occupation probability of the system in
its $m$-th energy level, while the vertical axis indicates the eigenenergy
of the $i$th energy level of the system. The solid orange curve represents
the isothermal process, which follows $p_{m}=e^{-\beta E_{m}}/\sum_{m}e^{-\beta E_{m}}$.
The red vertical lines and the blue horizontal dashed lines represent
the quantum adiabatic and quantum isochoric processes, respectively. }
\end{figure}

The isothermal process is an ideal process based on the quasi-static
assumption that the changing speed of the system's energy levels is
far slower than the relaxation of the system contacting with the heat
bath, so that the system is constantly on the thermal equilibrium
state with the same temperature of the heat bath. However, in the
finite time quasi-isothermal process, the system deviates from the
thermal equilibrium state. In our model, we consider the finite-time
process, where the system is not necessary in thermal equilibrium
for all the time.

The system of interest, initially in the thermal equilibrium state
with inverse temperature $\beta=(k_{B}T)^{-1}$, has $M$ discrete
energy levels $\{E_{m}^{(0)}\}$ ($m=1,2...,M$). And the corresponding
occupation probabilities in these energy levels are $\{p_{m}^{(0)}\}$
with $p_{m}^{(0)}=e^{-\beta E_{m}^{(0)}}/\sum_{m}e^{-\beta E_{m}^{(0)}}$.
To approach the quasi-isothermal process, we introduce the discrete
isothermal process (DIP) \cite{discrete-isothermal-process,DIP},
which includes a series of quantum adiabatic processes and quantum
isochoric processes \cite{QIP1,QIP2}.

As illustrated in Fig. \ref{fig-discrete isothermal process-1}, the
ideal isothermal process (orange solid curve) is decomposed by a series
of $N$ alternating short quantum adiabatic processes (red solid lines)
and quantum isochoric processes (blue dashed lines). One big difference
from the ideal isothermal process is that the discrete quasi-isothemal
process is operated within a finite duration $t_{f}$. We assume the
$j-$th sub quantum isochoric process takes time $\tau_{j}$ ($j=1,2,\dots,N$),
while each sub quantum adiabatic process is operated by a sudden quench
with no time cost. We assume that all the instantaneous eigen levels
$E_{m}$ are always avoided level-crossing. Therefore, even though
the adiabatic process is quenched, the instantaneous eigen states
keep unchanged and the quantum adiabatic condition is satisfied \cite{QIP2,Messiah_adiabatic}.

The operation time of the $j-$th step $\tau_{j}$ can not be infinitesimal
due to the requirement of the thermalization process. Detailed discussion
about the time scale of the step time will appear later. Through the
whole quasi-isothermal process, the energy spectra of the system are
changed from $\{E_{m}^{(0)}\}$ to $\{E_{m}^{(N)}\}$, while the corresponding
occupation probabilities turn from $\{p_{m}^{(0)}\}$ to $\{p_{m}^{(N)}\}$.
In the $j$-th subprocess, the $m$-th energy level is pulled from
$E_{m}^{(j-1)}$ to $E_{m}^{(j)}=E_{m}^{(j-1)}+\epsilon{}_{m}^{(j)}$
in the quantum adiabatic process (the occupation possibility is not
changed in this process). The tuning scheme of the energy levels $\{\epsilon_{m}^{(j)}\}$
can be described by a control function $f_{m}(j)=\sum_{k=1}^{j}\epsilon{}_{m}^{(k)}$,
with constraints $f_{m}(N)=E_{m}^{(N)}-E_{m}^{(0)}\equiv\Delta_{m}$
and $f_{m}\left(0\right)=0$. And the time to reach the $j-th$ step
is $t_{j}=\sum_{k=1}^{j}\tau_{k}$. Each sub quantum adiabatic process
is followed by a sub quantum isochoric process, during which the corresponding
occupation possibility is changed from $p_{m}^{(j-1)}$ to $p_{m}^{(j)}$
without shifting energy levels. The occupation possibility then is
assumed to relax to the corresponding equilibrium state with probability

\begin{equation}
p_{m}^{(j)}=\frac{e^{-\beta E_{m}^{(j)}}}{\sum_{m=1}^{M}e^{-\beta E_{m}^{(j)}}},\label{eq:possible}
\end{equation}
noticing the step time $\tau_{j}$ should be far larger than the typical
relaxation time of the heat bath to fulfill the low-dissipation condition.
The deviation from the equilibrium state is explicitly evaluated in
the Appendix A with an example of two-level atom. In the $j$-th subprocess,
there is no heat exchange between the system and bath in adiabatic
process, so the heat transfer only appears in the isochoric process
\cite{QIP2,heat-and-work} with $\Delta Q^{(j)}=\sum_{m=1}^{M}E_{m}^{(j)}\delta p_{m}^{(j)}$
, where $\delta p_{m}^{(j)}\equiv p_{m}^{(j)}-p_{m}^{(j-1)}$. Thus,
the heat transfer in the whole process $\Delta Q=\sum_{j=1}^{N}\Delta Q^{(j)}$
can be explicitly written as

\begin{equation}
\Delta Q=\sum_{j=1}^{N}\sum_{m=1}^{M}[E_{m}^{(0)}+\sum_{k=1}^{j}\epsilon_{m}^{(k)}]\delta p_{m}^{(j)}.\label{eq:Qn-1}
\end{equation}
In the high temperature limit $\beta E_{m}\ll1$, by keeping the first
order of $\beta E_{m}$, the above equation is simplified as 

\begin{equation}
\Delta Q=\sum_{j=1}^{N}\sum_{m=1}^{M}[-\frac{\beta}{M}\epsilon_{m}^{(j)}+\frac{\beta}{M^{2}}\sum_{m=1}^{M}\epsilon_{m}^{(j)}][E_{m}^{(0)}+\sum_{k=1}^{j}\epsilon{}_{m}^{(k)}].\label{eq:anyQ}
\end{equation}
On the other hand, the change in Shannon entropy $S=-\sum_{m=1}^{M}p_{m}\ln p_{m}$
($k_{B}=1$) only depends on the initial and final state of the system,
namely,

\begin{align}
\Delta S & =S^{(N)}-S^{(0)}\nonumber \\
 & =-\frac{\beta^{2}\sum_{m=1}^{M}(E_{m}^{(N)})^{2}}{M}+\frac{\beta^{2}(\sum_{m=1}^{M}E_{m}^{(N)})^{2}}{2M^{2}}\nonumber \\
 & +\frac{\beta^{2}\sum_{m=1}^{M}(E_{m}^{(0)})^{2}}{M}-\frac{\beta^{2}(\sum_{m=1}^{M}E_{m}^{(0)})^{2}}{2M^{2}}.\label{eq:deltaS}
\end{align}
With the heat exchange in Eq. (\ref{eq:deltaS}) and the entropy change
in Eq. (\ref{eq:anyQ}), we obtain the irreversible entropy generation
$S^{(\mathrm{ir})}=\Delta S-\Delta Q/T$ as 

\begin{equation}
S^{(\mathrm{ir})}=\frac{\beta^{2}}{2M}\sum_{j=1}^{N}\left[\sum_{m=1}^{M}(\epsilon_{m}^{(j)})^{2}-\frac{1}{M}(\sum_{m=1}^{M}\epsilon_{m}^{(j)})^{2}\right].\label{eq:S(ir)-e}
\end{equation}
For simplicity, we consider the case that the operation time of each
subprocess is the same, namely, $\tau_{j}=\tau$. And the total operation
time is $t_{f}=N\tau$. Then, by introducing the average tuning speed
of each step $v_{m}^{(j)}\equiv\epsilon_{m}^{(j)}/\tau$, we simplify
Eq. (\ref{eq:S(ir)-e}) as

\begin{equation}
S^{(\mathrm{ir})}=\frac{\beta^{2}\overline{\Delta}^{2}}{2(t_{f}/\tau)}\frac{\left\langle \overline{v^{2}}\right\rangle -\left\langle \overline{v}^{2}\right\rangle }{\left\langle \overline{v}\right\rangle ^{2}},\label{eq:main result}
\end{equation}
where \begin{subequations} 
\begin{align}
 & \left\langle \overline{v}\right\rangle \equiv\frac{1}{N}\sum_{j=1}^{N}[\frac{1}{M}\sum_{m=1}^{M}v_{m}^{(j)}],\label{appa-1}\\
 & \left\langle \overline{v^{2}}\right\rangle \equiv\frac{1}{N}\sum_{j=1}^{N}[\frac{1}{M}\sum_{m=1}^{M}\left(v_{m}^{(j)}\right)^{2}],\\
 & \left\langle \overline{v}^{2}\right\rangle \equiv\frac{1}{N}\sum_{j=1}^{N}\left(\frac{1}{M}\sum_{m=1}^{M}v_{m}^{(j)}\right)^{2}.
\end{align}
\end{subequations} Here $\overline{\bullet}$ means the average over
the energy levels, while $\left\langle \bullet\right\rangle $ means
the average over the whole process. $\overline{\Delta}=\sum_{m=1}^{M}\Delta_{m}/M$
is the average energy difference of the system's energy levels. Eq.
(\ref{eq:main result}) is the main result of this paper, and its
importance lies in two aspects. Firstly, Eq. (\ref{eq:main result})
shows that the irreversible entropy generation follows the $\mathscr{C}/t$
relation.With the current result, we basically answer the first question
posted in the introduction.

Secondly, the result in Eq. (\ref{eq:main result}) shows the explicit
dependence of the coefficient $\mathscr{C}$ on the control scheme
via the fluctuation of tuning speed. To further decouple the system
constants and the control scheme, we define

\begin{equation}
\Theta\equiv\frac{\beta^{2}\overline{\Delta}^{2}}{2},\xi\equiv\frac{\left\langle \overline{v^{2}}\right\rangle -\left\langle \overline{v}^{2}\right\rangle }{\left\langle \overline{v}\right\rangle ^{2}},\label{eq:definition}
\end{equation}
and rewrite the irreversible entropy generation as

\begin{equation}
S^{(\mathrm{ir})}=\frac{\Theta\xi}{t_{f}/\tau}.\label{eq:Sir}
\end{equation}
Here, $\Theta$ is related to the starting and ending point of the
stepwise isothermal process, $\xi$ shows the impact of different
control scheme. The coefficient is $\mathscr{C}=\Theta\xi\tau$. The
above relation clearly shows that the irreversible entropy generation
$S^{(ir)}\rightarrow0$ in the limit $t_{f}\rightarrow\infty$($N\rightarrow\infty$),
which is consistent with the Quasi-static isothermal process. 

When we consider the total operation time of the discrete isothermal
process (DIP), we have two adjustable parameters, namely the step
operation time $\tau$ and the total step number $N$. So the total
operation time $t_{f}$ can also be increased by increasing the step
time $\tau$. However, the irreversible entropy generation $S^{(ir)}$
approaches a fixed non-zero value, when increase the total operation
time $t_{f}$ via increasing the step time $\tau$ with fixed step
number $N$. In such case, the DIP will not back to the isothermal
process, and thus the requirement of recovering the Carnot cycle in
the limit $t_{f}\rightarrow\infty$ will not be fulfilled. Therefore,
in our derivation, we fix the step time and choose the total step
number $N$ to be the adjustable parameter.

\begin{table}
\begin{centering}
\begin{tabular}{c|c|c}
\hline 
\noalign{\vskip3pt}
$f(k)$  & $\xi$  & $S^{(\mathrm{ir})}$\tabularnewline[3pt]
\hline 
\noalign{\vskip3pt}
$ak^{n}$  & $\frac{n^{2}}{2n-1}$  & $\frac{n^{2}}{2n-1}\frac{\beta^{2}\Delta^{2}}{8N}$\tabularnewline[3pt]
\hline 
\noalign{\vskip3pt}
$b\left(\textrm{e}^{ak}-1\right)$  & $\left(\frac{1}{2}+\frac{b}{\Delta}\right)\ln(\frac{\Delta}{b}+1)$  & $\left(\frac{1}{2}+\frac{b}{\Delta}\right)\ln(\frac{\Delta}{b}+1)\frac{\beta^{2}\Delta^{2}}{8N}$\tabularnewline[3pt]
\hline 
\noalign{\vskip3pt}
$a\ln\left(bk+1\right)$  & $\frac{\sinh^{2}\left(\Delta/2a\right)}{\left(\Delta/2a\right)^{2}}$  & $\frac{\sinh^{2}\left(\Delta/2a\right)}{\left(\Delta/2a\right)^{2}}\frac{\beta^{2}\Delta^{2}}{8N}$\tabularnewline[3pt]
\hline 
\noalign{\vskip2pt}
\end{tabular}
\par\end{centering}
\caption{Irreversible entropy generation of different typical control functions
for the case of two-level atom. Here, each function satisfies the
constraint $f(0)=0$ and $f(N)=\Delta$, where $\Delta$ is the energy
level change during the DIP. In the calculation, we have already assumed
that $N\gg1$}

\label{tab:ft}
\end{table}

In the case of the two-level system ($M=2$), whose ground state energy
$E_{1}$ is taken as 0 in the whole process, Eq. (\ref{eq:main result})
reduces to

\begin{equation}
S^{(\mathrm{ir})}=\frac{\beta^{2}\Delta^{2}}{8t_{f}/\tau}\frac{\left\langle v^{2}\right\rangle }{\left\langle v\right\rangle ^{2}},\label{eq:Q-S}
\end{equation}
where $\left\langle v\right\rangle =\sum_{j=1}^{N}v_{2}^{(j)}/N$
and $\left\langle v^{2}\right\rangle =\sum_{j=1}^{N}\left(v_{2}^{(j)}\right)^{2}/N$
are the average of tuning speed and average of the square of tuning
speed, respectively. $\Delta=E_{2}^{(N)}-E_{2}^{(0)}$ is the energy
change of the excited state of the two-level system. When the energy
level control function $f(k)$ is linear dependent on the step, $\xi$
reaches the minimal value 1, in which case the the irreversible entropy
generation takes the minimal value, namely, $S^{(ir)}=\beta^{2}\Delta^{2}/(8t_{f}/\tau).$
This result shows that the irreversible behavior of the system can
be effectively reduced by optimizing the control protocol of the system's
energy levels in the DIP. A similar idea was reported in the optimization
of quantum Otto heat engine \cite{super-adibatic}, where the authors
introduced the super-adiabatic process to achieve zero friction in
the thermodynamic cycle. To make sure the system is at thermal equilibrium
in the end of each subprocess, the step time $\tau$ should be larger
than the relaxation time $1/\widetilde{\gamma}$, namely, $\tau>1/\widetilde{\gamma}$.
Here, $\widetilde{\gamma}=2\gamma/\left(\beta E^{(0)}\right)$ is
obtained in the high temperature limit (see Appendix A), and $\gamma$
is the system-bath coupling constant. For the nonlinear control functions
of time, i.e. $\xi>1$, the corresponding irreversible entropy generation
is larger than that of the linear case.

In Table. \ref{tab:ft}, we demonstrate the exact expressions of the
irreversible entropy generation related to three typical control functions
with the two-level atom example. When the control function is taken
as power function, i.e., $f(k)\propto k{}^{n}$, the irreversible
entropy generation follows a simple relation as $S_{n}^{(\mathrm{ir})}=n^{2}S_{1}^{(\mathrm{ir})}/(2n-1)$,
where $S_{1}^{(\mathrm{ir})}=\beta^{2}\Delta^{2}/(8t_{f}/\tau)$.
This relation is confirmed by the master equation based numerical
results (the points) as illustrated in Fig. \ref{fig:Irreversible},
where we plot the irreversible entropy generation as a function of
operation time with $N=t_{f}/\tau\in[20,120]$. In the simulation,
we fix the step time with $\tau=1$ and increase the number of steps
$N$. The initial energy of excited state is $E_{2}^{\left(0\right)}=10$
and the final one is $E_{2}^{\left(N\right)}=6$. The inverse temperature
is $\beta=0.1$, and the decay rate is $\gamma=1$. The adiabatic
process is assumed to be instantaneous. During the isochoric process,
the evolution of the system is govern by the master equation as shown
in Eq. (\ref{eq:master equation}). With the increase of $n$ of the
control function $f(k)$, the irreversible entropy generation is also
increased as illustrated in Fig. \ref{fig:Irreversible}. The numerical
results are in good agreement with theoretical prediction in Eq. (\ref{eq:Q-S}).
\begin{flushleft}
\begin{figure}
\includegraphics[width=8.5cm]{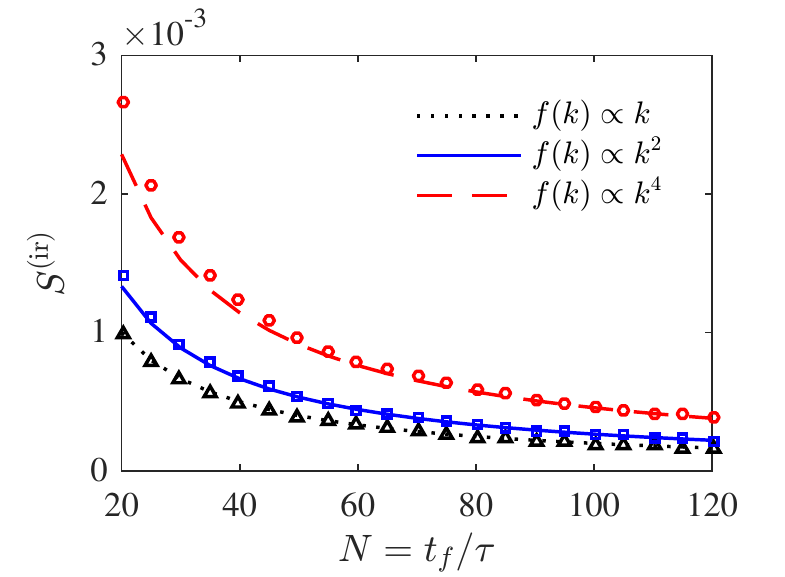}\caption{\label{fig:Irreversible}(Color online) Irreversible entropy generation
as function of operation time, where $f(k)=\Delta(k/N)$ (black),
$f(k)=\Delta(k/N)^{2}$ (blue), and $f(k)=\Delta(k/N)^{4}$ (red).
The points represent the numerical results, and the lines represent
the analytical results of Eq. (\ref{eq:Sir}). Here $\beta=0.1$ is
the inverse temperature of the heat bath, $E^{(0)}=10$ is the initial
energy of the two-level system's excited state, $\Delta=4$ is the
energy change of the excited state, and $\gamma=1$ is the system-bath
coupling strength.}
\end{figure}
\par\end{flushleft}

With our main result in Eq.(\ref{eq:main result}), we basically answer
the two questions: (1) the $\mathscr{C}/t$ relation is valid at least
in our discrete isothermal process, (2) the irreversible entropy generation
coefficient $\mathscr{C}$ is proportional to the variance of the
tuning speed. This result allows us to design optimal heat engine
cycle.

\section{Efficiency of a Carnot-like heat engine}

In this section, we will construct a quantum Carnot-like (QCL) heat
engine to demonstrate the concrete control scheme of achieving the
EMP. The isothermal processes in the normal Carnot cycle will be replaced
with our discrete isothermal processes.

\begin{figure}
\begin{centering}
\includegraphics[width=8.5cm]{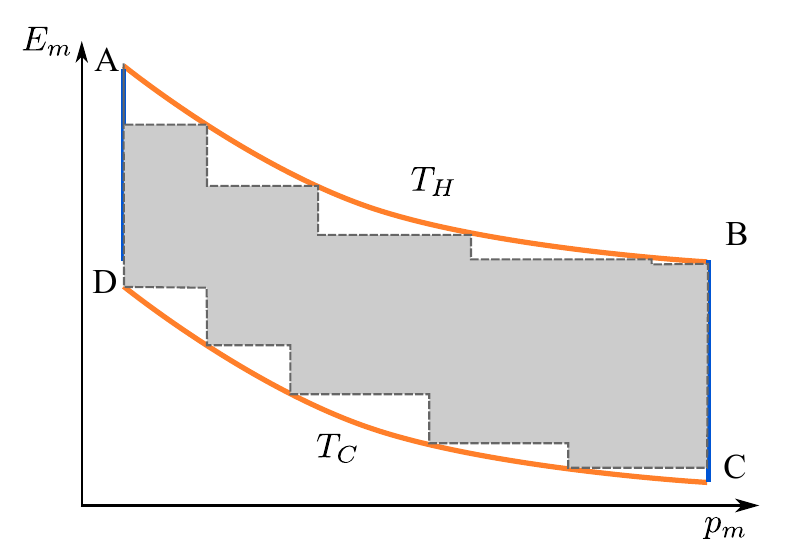} 
\par\end{centering}
\caption{\label{fig-disreate carnot cycle }(Color online) Schematic illustration
of a discrete Carnot cycle. This discrete Carnot-like thermodynamic
cycle is composed by two discrete isothermal processes ($\textrm{A}\rightarrow\textrm{B}$
and $\textrm{C}\rightarrow\textrm{D}$) and two adiabatic processes
($\textrm{B}\rightarrow\textrm{C}$ and $\textrm{D}\rightarrow\textrm{A}$)
. Here the discrete black dashed curve represents the discrete isothermal
process, while the smooth orange solid curve represents the corresponding
quasi-static isothermal process in the limit that the operation time
approaches infinity. The gray area encircled by the two discrete curve
represents the output work per cycle. The temperature of the high
and low temperature bath are $T_{H}$ and $T_{C}$, respectively,
and the operation time that the working substance contacts with them
are $t_{H}$ and $t_{C}$, respectively.}
\end{figure}

With the well-defined DIP, we construct the discrete Carnot-like thermodynamic
cycle, as illustrated in Fig. \ref{fig-disreate carnot cycle }, with
two discrete isothermal processes ($\textrm{A}\rightarrow\textrm{B}$
and $\textrm{C}\rightarrow\textrm{D}$) and two adiabatic processes
($\textrm{B}\rightarrow\textrm{C}$ and $\textrm{D}\rightarrow\textrm{A}$).
The two discrete isothermal processes are realized by contacting two
heat baths with temperature $T_{H}$ and $T_{C}$, respectively. Without
losing generality, we consider the simplest case with the two level
system as the working substance to clearly show the design scheme.
For the two-level system, in each cycle, the energy of its ground
state $\left|g\right\rangle $ is fixed at 0, while the energy level
of the excited state $\left|e\right\rangle $ is tuned by an outsider
agent to extract work, namely $H=E\left(t\right)\left|e\right\rangle \left\langle e\right|$.
To optimize the heat engine, we consider two different control functions
for the DIPs, namely, $E\left(t\right)=E_{H}^{(0)}+f_{H}(t)\,\textrm{(A}\rightarrow\textrm{B)}$
and $E\left(t\right)=E_{C}^{(0)}+f_{C}(t)\,\textrm{(C}\rightarrow\textrm{D)}$.
Here, $E_{H}^{(0)}$ ($E_{C}^{(0)}$) is the initial energy of the
excited state in the high (low) temperature DIP, and $f_{H}(t)$ ($f_{C}(t)$)
is the corresponding control function. Noticing that we have the constraint
for the control functions $f_{H}(t_{H})=E_{H}^{(N)}-E_{H}^{(0)}$
and $f_{C}(t_{C})=E_{C}^{(N)}-E_{C}^{(0)}$, where $t_{H}$ ($t_{C}$)
is the operation time of the two DIPs and$E_{H}^{(N)}$ ($E_{C}^{(N)}$)
is the corresponding finial energy of the excited state.

The heat transfer of the QCL heat engine is written as $\sum_{j=1}^{N}E^{(j)}\delta p^{(j)}$
\cite{QIP2,heat-and-work}, we can connect the heat absorb (release)
from the heat bath $\Delta Q_{H}$ ($\Delta Q_{C}$) to the area $\Lambda_{H}$
($\Lambda_{C}$) encircled by the high (low) temperature-related discrete
black dashed curve and the horizon axis. Thus, the power and efficiency
of such a QCL heat engine can be expressed by two characteristic areas
as follows

\begin{equation}
P=\frac{W}{t_{H}+t_{C}}=\frac{\Lambda_{H}-\Lambda_{C}}{t_{H}+t_{C}}=\frac{\Delta\Lambda}{t_{H}+t_{C}},\label{eq:power}
\end{equation}

\begin{equation}
\eta=\frac{W}{\Delta Q_{H}}=\frac{\Lambda_{H}-\Lambda_{C}}{\Lambda_{H}}=1-\frac{\Lambda_{C}}{\Lambda_{H}}.\label{eq:eff}
\end{equation}
Here, $\Delta\Lambda=\Lambda_{H}-\Lambda_{C}$ corresponds to the
output work per cycle as represented by the gray area in Fig. \ref{fig-disreate carnot cycle }.
We have assumed the energy level is tuned very rapid in the two adiabatic
processes ($\textrm{B}\rightarrow\textrm{C}$ and $\textrm{D}\rightarrow\textrm{A}$),
so that the corresponding operation time is ignored. It can be seen
from Fig. \ref{fig-disreate carnot cycle } that the area $\Lambda_{H}$
is smaller than the area $\Lambda_{H}^{(r)}$ encircled by the high
temperature-related smooth orange solid curve and the horizon axis,
i.e., $\Lambda_{H}<\Lambda_{H}^{(r)}$. While the area $\Lambda_{C}$
is lager than the area $\Lambda_{C}^{(r)}$ encircled by the low temperature-related
smooth orange solid curve and the horizon axis, i.e., $\Lambda_{C}>\Lambda_{C}^{(r)}$.
Since $\Lambda_{H}^{(r)}$ and $\Lambda_{C}^{(r)}$ are connected
to the reversible heat absorb and reversible heat release, respectively,
the following inequality can be easily verified

\begin{equation}
\eta=1-\frac{\Lambda_{C}}{\Lambda_{H}}<1-\frac{\Lambda_{C}^{(r)}}{\Lambda_{H}^{(r)}}=\eta_{C}.
\end{equation}
Following from Eq. (\ref{eq:Sir}), we obtain $\Lambda_{H}=T_{H}\Delta S_{H}-\xi_{H}\Theta_{H}/t_{H}$
and $\Lambda_{C}=T_{C}\Delta S_{C}-\xi_{C}\Theta_{C}/t_{C}$, where
$\Delta S_{C}=\Delta S_{H}$ corresponds to the reversible part of
the heat transfer in the whole cycle. Noticing the efficiency in Eq.
(\ref{eq:eff}) and the power in Eq. (\ref{eq:power}) are now connected
with each other through the operation time $t_{C}$ and $t_{H}$.
Thus, to find the EMP of the heat engine, one should optimize the
power via the two operation times. With the framework developed by
Esposito et. al. \cite{low-dissipation}, the EMP of such heat engine
reads

\begin{equation}
\eta_{\textrm{EMP}}=\frac{\eta_{C}\left(1+\sqrt{\frac{T_{C}\xi_{C}\Theta_{C}}{T_{H}\xi_{H}\Theta_{H}}}\right)}{\left(1+\sqrt{\frac{T_{C}\xi_{C}\Theta_{C}}{T_{H}\xi_{H}\Theta_{H}}}\right)^{2}+\frac{T_{C}}{T_{H}}\left(1-\frac{\xi_{C}\Theta_{C}}{\xi_{H}\Theta_{H}}\right)},
\end{equation}
where we have replaced the phenomenological parameter $\Sigma$ in
the original result of Ref. \cite{low-dissipation} by our system
and control scheme related parameters $\xi$ and $\Theta$, namely,
$\Sigma_{\alpha}\rightarrow\xi_{\alpha}\Theta_{\alpha}$ ($\alpha=C,H$).
Using the definition of $\Theta$ in Eq. (\ref{eq:definition}) we
have the ratio of $\Theta_{C}$ and $\Theta_{H}$ explicitly expressed
as $\Theta_{C}/\Theta_{H}=\beta_{C}^{3}\Delta_{C}^{2}E_{C}^{(0)}\gamma_{C}^{-1}/\left(\beta_{H}^{3}\Delta_{H}^{2}E_{H}^{(0)}\gamma_{H}^{-1}\right)=\gamma_{H}/\gamma_{C}$.
Here, we fix the four points A, B, C and D as the same as that in
normal Carnot cycle with the relations $\beta_{H}E_{H}^{(0)}=\beta_{C}E_{C}^{(0)}$
and $\beta_{H}E_{H}^{(N)}=\beta_{C}E_{C}^{(N)}$. The operation time
of each sub-process of the DIP is taken as $\tau_{\alpha}=2\widetilde{\gamma}_{\alpha}^{-1}=\beta_{\alpha}E_{\alpha}^{(0)}/\gamma_{\alpha}$
($\alpha=C,H$). For a practical designed heat engine, we typically
have $\gamma_{H}/\gamma_{C}$ fixed once the interaction between the
system and the heat bath is given. Therefore, the EMP of such a heat
engine only depends on the scheme that the working substance's energy
spectra being tuned, i.e., $\xi_{C}/\xi_{H}$. When the control function
is exponential function, we choose the system's energy spectra to
be linearly tuned at the low temperature bath, that is, $n_{C}=1$,
and the control function satisfies $f_{H}(t)\propto t^{n_{H}},n_{H}\gg1$
at the high temperature bath. Then, the EMP is simplified as

\begin{equation}
\eta_{\textrm{EMP}}=\frac{\eta_{C}\left(1+\sqrt{\frac{2T_{C}}{n_{H}T_{H}}}\right)}{2-\eta_{C}+2\sqrt{\frac{2T_{C}}{n_{H}T_{H}}}}\approx\eta_{+}\left(1-\eta_{+}\sqrt{\frac{2(1-\eta_{C})}{n_{H}}}\right),\label{eq:eta_EMP}
\end{equation}
where $\eta_{+}=\eta_{C}/(2-\eta_{C})$.

\begin{figure}
\begin{centering}
\includegraphics[width=8.5cm]{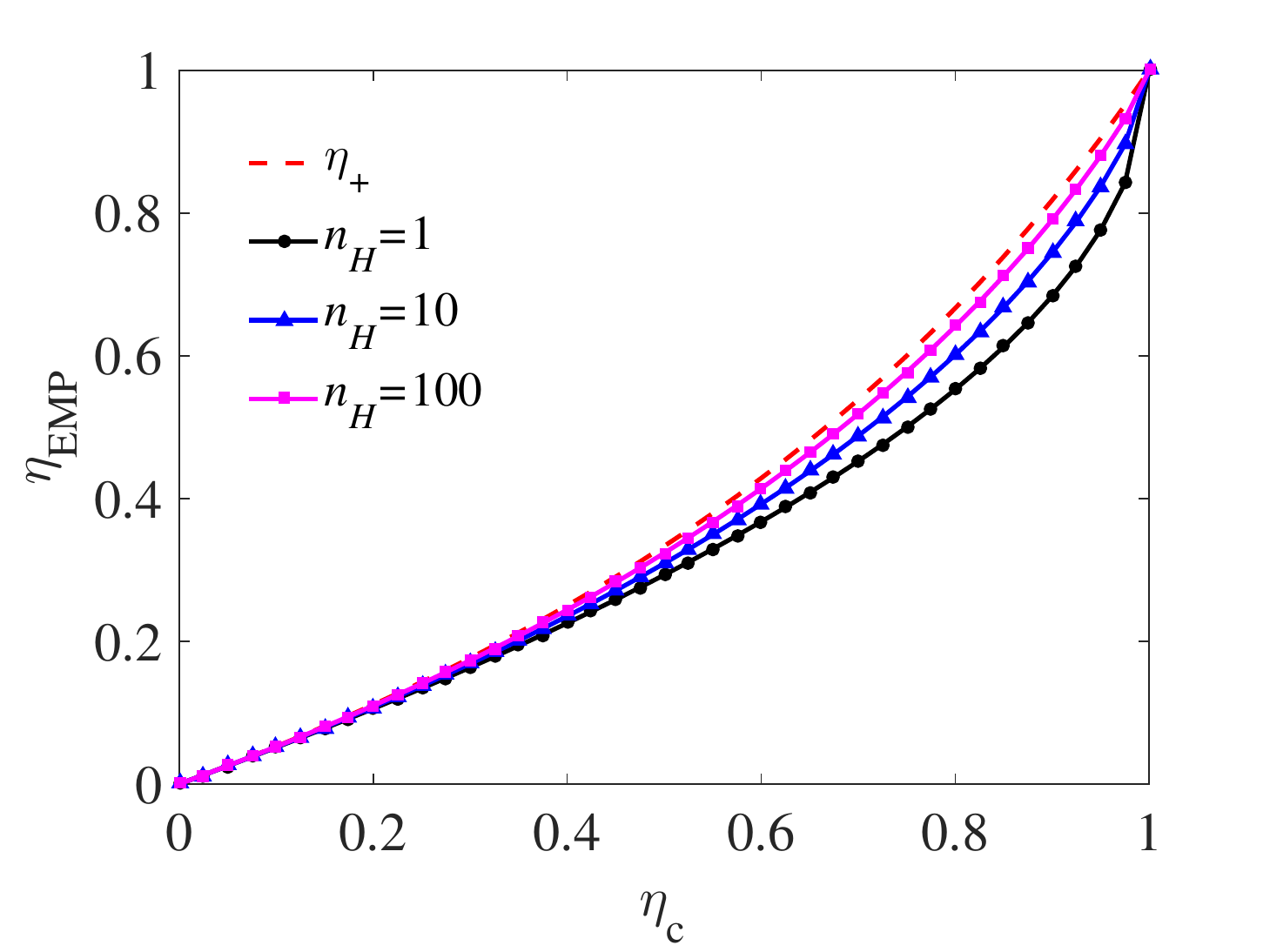} 
\par\end{centering}
\caption{\label{fig:EMP}Efficiency at maximum power as function of Carnot
efficiency, where the red dashed curve indicates the maximum value
that the EMP can achieve $\eta_{+}=\eta_{C}/(2-\eta_{C})$. The black
circle line represents the Curzon-Ahlborn efficiency with $n_{H}=n_{C}=1$,
namely $f_{H}\propto t$. The blue triangle line and pink square line
correspond to the control function in the high temperature discrete
isothermal process being taken as $f_{H}\propto t^{10}$ and $f_{H}\propto t^{100}$,
respectively.}
\end{figure}

In Fig. \ref{fig:EMP}, we show the achievable EMP for different control
functions. The dashed red line shows the maximum value $\eta_{+}$
of the EMP. And the black circle line represents the Curzon-Ahlborn
efficiency $\eta_{\textrm{CA}}=1-\sqrt{1-\eta_{C}}$, which can be
realized in our scheme with $n_{H}=n_{C}=1$. It can be seen from
Eq. (\ref{eq:eta_EMP}) that the EMP of the heat engine can be adjusted
via different control functions, and can be significantly improved
with $n_{H}$ increasing. The controllability of EMP is also demonstrated
in Fig. \ref{fig:EMP} through the exact numerical results. With the
increase of $n_{H}$, the EMP is deviating from $\eta_{\textrm{CA}}$
and getting closer to the upper bound of the EMP $\eta_{+}$. This
means that it is feasible to control the EMP of the heat engines through
different control schemes of tuning the system's energy levels in
the DIPs. Even with the constraint relation between power and efficiency
\cite{TLA-heat-engine,constriant2-1,constriant2}, we notice that
one can maintain the maximum output power while increasing the EMP
via different control schemes. Detailed discussion is shown in Appendix
B. 

The current control scheme with the stepwise Carnot-like cycle makes
it possible for the experimental test by the widely-used setups \cite{experiment1,experiment2,experiment-trapped-ion}
for testing Jarzynski equality. Experiments concerning EMP were not
feasible, to our best knowledge, because of unavailability of the
control scheme. Our stepwise control scheme allows a clear separation
of heat exchange and work extraction processes for implementing measurement.
For the clarity, we consider the simple two-level atom case. In the
DIP ($A\rightarrow B)$, we measure the probability sequences $p_{e}^{\left(0\right)}\rightarrow p_{e}^{\left(1\right)}\rightarrow p_{e}^{\left(2\right)}\rightarrow\cdots\rightarrow p_{e}^{\left(N-1\right)}\rightarrow p_{e}^{\left(N\right)}$
along with the energy level changes $E^{\left(0\right)}\rightarrow E^{\left(0\right)}\rightarrow E^{\left(2\right)}\rightarrow\cdots\rightarrow E^{\left(N-1\right)}\rightarrow E^{\left(N\right)}$.
The heat absorbed is calculated as follow, 
\begin{equation}
\Delta Q_{H}=\sum_{i=0}^{N-1}(p_{e}^{\left(i+1\right)}-p_{e}^{\left(i\right)})E^{\left(i+1\right)}.
\end{equation}
The first test is the $\mathscr{C}/t$ relation in Eq. (\ref{eq:Q-S})
with variation of the operation time $t_{f}$ as well as different
control schemes listed in Table \ref{tab:ft}. The irreversible entropy
generation $S^{\left(\mathrm{ir}\right)}$ is obtained as 
\begin{equation}
S^{\left(\mathrm{ir}\right)}=\Delta S-\frac{\Delta Q_{H}}{T_{H}},
\end{equation}
where $\Delta S=p_{e}^{\left(N\right)}\ln p_{e}^{\left(N\right)}+[1-p_{e}^{\left(N\right)}]\ln[1-p_{e}^{\left(N\right)}]-p_{e}^{\left(0\right)}\ln p_{e}^{\left(0\right)}+[1-p_{e}^{\left(0\right)}]\ln[1-p_{e}^{\left(0\right)}]$.
The similar approach is applied for the DIP ($C\rightarrow D$) for
the heat $\Delta Q_{C}$ directed to the low temperature bath. The
power of the engine is obtained as $P=(\Delta Q_{H}-\Delta Q_{C})/(t_{H}+t_{C})$.
To meet the requirement of operation time for the heat engine achieving
the EMP \cite{low-dissipation}, $t_{H}$ and $t_{C}$ follow $t_{H}=2\xi_{H}\Theta_{H}\left[1+\sqrt{T_{C}\xi_{C}\Theta_{C}/\left(T_{H}\xi_{H}\Theta_{H}\right)}\right]/\left(\eta_{C}\Delta S\right)$
and $t_{C}=t_{H}\sqrt{T_{C}\xi_{C}\Theta_{C}/\left(T_{H}\xi_{H}\Theta_{H}\right)}$,
respectively. Here, $\xi_{\alpha}$ is determined by the specific
form of the control function $f_{\alpha}(t)$ ($\alpha=C,H$) as demonstrated
in Table. \ref{tab:ft}. $\Theta_{H}$ ($\Theta_{C}$) is related
to the starting and ending point of the high (low) temperature DIP
as given by Eq. (\ref{eq:definition}). The efficiency for one specific
control scheme is obtained as $\eta=(\Delta Q_{H}-\Delta Q_{C})/\Delta Q_{H}$.

\section{Conclusion}

In summary, by introducing the discrete isothermal process, we presented
a general proof of the inverse relation between the irreversible entropy
generation and time in finite time isothermal process, namely $S^{(\mathrm{ir})}=\mathscr{C}/t$,
which is widely used for the actual heat engines within the low-dissipation
region. Besides the system constants, we showed that the coefficient
$\mathscr{C}$ of irreversible entropy generation also depends the
control scheme when the system's energy levels are tuned in the discrete
isothermal process. Remarkably, the minimal irreversible entropy generation
is achieved when the energy levels of the system are linearly tuned.
This discovery allows us to design optimal heat engine cycle. With
a two-level atomic heat engine as an illustration, we demonstrate
that the EMP of the heat engine can be optimized by applying different
control schemes when the working substance contacting with different
heat baths. The controllability of EMP can be experimentally verified
with some state of art experimental platforms, such as superconducting
circuit system \cite{Pekola2009}, and trapped ion \cite{experiment1,experiment-trapped-ion}. 
\begin{acknowledgments}
Y. H. Ma is grateful to H. Yuan for the helpful discussion. We thank
Z. C. Tu for the careful reading of this manuscript. This work is
supported by NSFC (Grants No. 11705008, Grants No. 11774323, No. 11534002),
the National Basic Research Program of China (Grant No. 2016YFA0301201
\& No. 2014CB921403), the NSAF (Grant No. U1730449 \& No. U1530401),
and Beijing Institute of Technology Research Fund Program for Young
Scholars.
\end{acknowledgments}

\appendix

\section{Evolution of the two level system }
\begin{widetext}
The dynamics of the two-level atom, when it contacts with a heat bath
with inverse temperature $\beta=(k_{B}T)^{-1}$, is described by the
master equation

\begin{equation}
\frac{dp_{e}\left(t\right)}{dt}=-\gamma\left(2n_{\mathrm{th}}[E(t)]+1\right)p_{e}\left(t\right)+\gamma n_{\mathrm{th}}[E(t)],\label{eq:master equation}
\end{equation}
where $p_{e}\left(t\right)\equiv\langle e\vert\hat{\rho}\left(t\right)\vert e\rangle$
is the excited state population of the density matrix $\hat{\rho}\left(t\right)$,
$\gamma$ the system-bath coupling strength, and $n_{\mathrm{th}}[E(t)]=1/\left(\exp[\beta E\left(t\right)]-1\right)$
the mean occupation number of bath mode with frequency $E\left(t\right)$.
The solution of Eq. (\ref{eq:master equation}) reads

\begin{equation}
p_{e}(t)=\frac{n_{\mathrm{th}}[E(t)]}{1+2n_{\mathrm{th}}[E(t)]}(1-e^{-\gamma\left\{ 2n_{\mathrm{th}}[E(t)]+1\right\} t})+e^{-\gamma\left\{ 2n_{\mathrm{th}}[E(t)]+1\right\} t}p_{e}\left(0\right)\;,\label{eq:solution}
\end{equation}
applying which to the $j-$th step of the discrete isothermal process,
we obatin

\begin{equation}
p_{e}\left(\tau\right)=p_{e}^{(j)}(1-e^{-\frac{1+e^{-\beta E^{(j)}}}{e^{-\beta E^{(j)}}-1}\gamma\tau})+e^{-\frac{1+e^{-\beta E^{(j)}}}{e^{-\beta E^{(j)}}-1}\gamma\tau}p_{e}^{(j-1)}.
\end{equation}
In the high temperature limit, i.e. , $\beta E\ll1$, the above equation
can be simplified as

\begin{equation}
p_{e}\left(\tau\right)=p_{e}^{(j)}+e^{-\frac{2\gamma\tau}{\beta E^{(j)}}}\left(p_{e}^{(j-1)}-p_{e}^{(j)}\right).
\end{equation}
Choosing $\tau=\beta E^{(0)}/\gamma$, one finds $\exp\left[-2\gamma\tau/\left(\beta E^{(j)}\right)\right]\approx e^{-2}\ll1$,
then $p_{e}\left(\tau\right)\rightarrow p_{e}^{(j)}$. The order of
the error is about $e^{-2}$.
\end{widetext}

\section{Dependence of power on control scheme}

For low-dissipation heat engines, the maximum power in Ref. \cite{TLA-heat-engine,low-dissipation}
can be re-written with our notation as,

\begin{align}
P_{max} & =\frac{(\eta_{C}T_{h}\Delta S)^{2}}{4(\sqrt{T_{h}\xi_{h}\Theta_{h}}+\sqrt{T_{c}\xi_{c}\Theta_{c}})^{2}}\\
 & =\frac{(\eta_{C}T_{h}\Delta S)^{2}}{4T_{c}\xi_{c}\Theta_{c}}\left(1+\sqrt{\frac{T_{h}\xi_{h}\Theta_{h}}{T_{c}\xi_{c}\Theta_{c}}}\right)^{-2}.
\end{align}
It is clear that the maximum power depends on $\xi_{h}/\xi_{c}$ and
$\xi_{c}$, while the EMP only depends on $\xi_{h}/\xi_{c}$ as shown
in {[}Eq. (14){]} of our manuscript. Therefore we can maintain the
maximum power unchanged by fixing the value of

\begin{equation}
\xi_{c}^{-1}\left(1+\sqrt{\frac{T_{h}\xi_{h}\Theta_{h}}{T_{c}\xi_{c}\Theta_{c}}}\right)^{-2},
\end{equation}
while improving the efficiency by increasing the ratio $\xi_{h}/\xi_{c}$. 
\end{document}